\documentclass[fleqn,5p,sort&compress,times]{elsarticle}
\usepackage{graphicx}
\usepackage{hyperref}
\usepackage{color}
\usepackage{latexsym}
\usepackage{amssymb,stackrel}

\newcommand{\be}{\begin{eqnarray}}
\newcommand{\ee}{\end{eqnarray}}

\usepackage{tikz}

\newcommand{\ba}{\begin{eqnarray}}
\newcommand{\ea}{\end{eqnarray}}
\def\bs{\begin{subequations}}
\def\es{\end{subequations}}
\def\a{\alpha}

\def\de{\delta}
\def\g{\gamma}

\def\la{\lambda}

\def\om{\omega}

\def\s{\sigma}

\def\vp{\varphi}
\def\N{\nabla}

\def\cF{\mathcal{F}}
\def\cG{\mathcal{G}}

\def\cK{\mathcal{K}}
\def\cL{\mathcal{L}}

\def\p{\partial}

\def\B{\Box}
\newcommand{\Eq}[1]{(\ref{#1})}

\def\cob{\color{blue}}
\newcommand{\book}[5]{{#1}, #2, #3, #4, #5}
\newcommand{\books}[4]{{#1}, #2, #3, #4}
\newcommand{\oarX}[1]{\href{http://arxiv.org/abs/#1}{{\ttfamily\cob arXiv:#1}}}
\newcommand{\arX}[1]{\href{http://arxiv.org/abs/#1}{{\ttfamily\cob arXiv:#1}}}
\newcommand{\doin}[6]{\href{http://dx.doi.org/#1}{{\cob #2 #3 {#4} (#6) #5}}}
\newcommand{\doinn}[5]{\href{http://dx.doi.org/#1}{{\cob #2 {#3} (#5) #4}}}
\newcommand{\doij}[5]{\href{http://dx.doi.org/#1}{{\cob #2 #3 (#5) #4}}}

\newcommand{\ndoinn}[5]{\href{#1}{{\cob #2 {#3} (#5) #4}}}

\newcommand{\tia}[1]{{#1},}

\def\rme{e}
\def\rmd{d}
\def\rmi{i}

\begin{document}

\begin{frontmatter}

\title{Non-perturbative spectrum of non-local gravity}

\author{Gianluca Calcagni}
\ead{g.calcagni@csic.es}
\address{Instituto de Estructura de la Materia, CSIC, Serrano 121, 28006 Madrid, Spain}

\author{Leonardo Modesto}
\ead{lmodesto@sustc.edu.cn}
\address{Department of Physics, Southern University of Science and Technology, Shenzhen 518055, China} 

\author{Giuseppe Nardelli}
\ead{giuseppe.nardelli@unicatt.it}
\address{Dipartimento di Matematica e Fisica, Universit\`a Cattolica del Sacro Cuore, via Musei 41, 25121 Brescia, Italy}
\address{TIFPA -- INFN c/o Dipartimento di Fisica, Universit\`a di Trento, 38123 Povo (Trento), Italy}

\begin{abstract}
We investigate the non-perturbative degrees of freedom of a class of weakly non-local gravitational theories that have been proposed as an ultraviolet completion of general relativity. At the perturbative level, it is known that the degrees of freedom of non-local gravity are the same of the Einstein--Hilbert theory around any maximally symmetric spacetime. We prove that, at the non-perturbative level, the degrees of freedom are actually eight in four dimensions, contrary to what one might guess on the basis of the ``infinite number of derivatives'' present in the action. It is shown that six of these degrees of freedom do not propagate on Minkowski spacetime, but they might play a role at large scales on curved backgrounds. We also propose a criterion to select the form factor almost uniquely.
\end{abstract}

\end{frontmatter}


\section{Introduction} 

A quantum theory of gravity \cite{Ori09,Fousp,CQC} should be able to solve, or say something constructive about, some problems left open in general relativity, such as the singularity problem (there exist spacetime points where the laws of physics break down, as in the big bang at the beginning of the Universe or inside black holes), the cosmological constant problem (two thirds of the content of our patch of the cosmos is made of a ``dark energy'' component not adequately described by general relativity or particle physics), and the mystery surrounding the birth and first stage of development of the Universe (the actual origin of the inflaton is unknown). In recent years, a new perturbative quantum field theory of gravity has rapidly emerged as a promising and accessible framework where the gravitational force consistently obeys the laws of quantum mechanics and all infinities seem to be tamed \cite{Kra87,Kuz89,Tom97,Mod1,BGKM,CaMo2,MoRa1,TBM,MoRa2,To15b,KMRS,MoRa3}. This proposal adapts ordinary techniques of perturbative field theory to an action with non-local operators. Fulfilling initial expectations based on na\"ive power-counting arguments, the theory turns out to be unitary and super-renormalizable or finite \cite{MoRa1} at the quantum level thanks to the non-local nature of its dynamics. Causality is not violated in the usual eikonal limit \cite{GiMo} and one expects non-offensive microcausality violations at the non-locality scale. The theory may resolve the big bang \cite{BMS,cuta8,BKM1,KV3,BKMV,Kos13,CMNi,EKM,ll16} and black-hole singularities \cite{Fro16,BMR,KoM,CaMo3,MyPa,Ed18b}, and its cosmological solutions may unravel interesting bottom-up scenarios in the early universe (inflation) and at late times (dark energy).

Surprisingly, these encouraging features are accompanied by a number of appalling gaps of knowledge on basic questions on the classical theory, such as how to find solutions of the dynamics and whether they match the singularity-free geometries found when linearizing the equations of motion. The dynamics is usually solved with approximations or assumptions which do not give access to all admissible solutions \cite{cuta2}. Also, when considering non-linear interactions (gravity is as non-linear as it can be!) it becomes unclear how many initial conditions one must specify and how many degrees of freedom (d.o.f.) populate the spectrum of physical particles of the theory. The purpose of this Letter is to pave the way to fill these gaps and infer the total number of non-perturbative d.o.f.\ in minimal non-local gravity \cite{Mod1,CaMo2} (super-renormalizable in $D=4$ dimensions; finite in $D=5$). 

We can summarize our findings in three statements. (A) For the types of non-locality giving rise to a renormalizable theory, the number of field d.o.f.\ is finite and equal to $D(D-2)=8$ in four dimensions. Two of these d.o.f.\ correspond to the graviton, they propagate on flat spacetime at short distances, and are a familiar acquaintance in the non-local linearized dynamics. The other d.o.f.\ are a novelty because they emerge only when the fully non-linear non-local dynamics is considered. Since they are not visible in a perturbative treatment on Minkowski spacetime, these d.o.f.\ are \emph{non-perturbative}, typical of curved backgrounds, and, hence, may be important in the description of \emph{large-scale, long-range physics}, such as at astrophysical or cosmological scales. (B) Also the number of initial conditions to specify in order to find classical solutions is finite. This contributes to, or even settles, a seventy-year-old debate about whether non-local theories are predictive at all, due to the peculiarities of their problem of initial conditions. The answer is Yes, for the specific non-localities considered here. Knowing how to construct dynamical solutions makes a fundamental step in the understanding of the capabilities of the specific theory under examination, both at the classical and the quantum level. (C) The system can be recast as a set of finite-order differential equations, and hence the number of initial conditions is finite, only for two specific non-local form factors among those circulating in the literature. This constrains the ambiguity on the choice of form factors, i.e., of non-local theory.

The most important consequence of identifying new d.o.f.\ and of knowing how to formulate the initial-condition problem is that now one can, on one hand, construct nontrivial cosmological and black-hole classical solutions previously inaccessible with current methods and, on the other hand, build a robust, rigorous, and systematic set of phenomenological models that can be tested against experiments and observations, a vital task for any candidate theory of natural phenomena. Therefore, the results of the present work may be of interest for the applied mathematician, the quantum-gravity theoretician and phenomenologist, the particle-physics theoretician, the cosmologist, and the astrophysicist.

After a brief overview of the most prominent non-local gravities, we show that the number of initial conditions and d.o.f.\ is finite and we count them explicitly. The logic to follow is simple: (i) we write the non-local dynamics with infinitely many derivatives as a system of ``master equations'' which are second order in spacetime derivatives, hence the Cauchy problem is well defined; (ii) from this reformulation, we extract the number of initial conditions and the d.o.f. The proof is self-contained and may be skipped by the reader interested only in the physical consequences of the theory, which are discussed above and in the final section.


\section{Non-local dynamics} 


\subsection{Brief overview of non-local quantum gravity}

Consider the classical action
\be
S = \frac{1}{2\kappa^2}\int \rmd^D x \sqrt{-g} \left[R+G_{\mu\nu}\gamma(\Box) R^{\mu\nu} \right]\,,\label{action}
\ee
where $G_{\mu\nu}$ is the Einstein tensor and $\gamma(\Box)$ is a non-local form factor, an entire function of $\B$ having special asymptotic properties \cite{Mod1,MoRa1,MoRa2,To15b,Kuz89,ALS,Tom97}. It can be parametrized as
\be\label{gamma}
\gamma(\Box) = \frac{\rme^{{\rm H}(\Box)}-1}{\Box}\,,
\ee
where ${\rm H}(\B)$ depends on the dimensionless combination $l^2\B$ and $l$ is a fixed length scale. {The four principal theories are shown in Tab.~\ref{tab1}. In the first case (``pol'': asymptotically polynomial), $P(z)$ is a real positive polynomial of degree $n$ ($2n$ derivatives) with $P(0) = 0$, $\Gamma$ is Euler function, and $\gamma_{\rm E}$ is the Euler--Mascheroni constant. In the second case (``exp'') the form factor is asymptotically exponential.
\begin{table}
\centering
\caption{Form factors in non-local gravity.\label{tab1}}
\begin{tabular}{c|c|c|l}\hline\hline
${\rm H}(\B)$ & $P(\B)$ & $f(\om)$ & Form factor name\\\hline
\begin{tabular}{@{}c@{}}${\rm H}^{\rm pol}(\B):=\a\{\ln P(\B)$\\ $\quad+\Gamma[0,P(\B)]+\gamma_{\rm E}\}$ \end{tabular} & \begin{tabular}{@{}c@{}}$-l^2\B$ \\ $O(\B^n)$ \end{tabular} & $\rme^{-\om}$ & \begin{tabular}{@{}l@{}}Kuz'min \cite{Kuz89} \\ Tomboulis \cite{Tom97,Mod1}\end{tabular}\\\hline
${\rm H}^{\rm exp}(\B):=\a P(\B)$ & \begin{tabular}{@{}c@{}}$-l^2\B$ \\ $l^4\B^2$\end{tabular} & $1-\om$
& \begin{tabular}{@{}l@{}}stringy \cite{BMS,CaMo2} \\ Krasnikov \cite{Kra87}\end{tabular}\\\hline\hline
\end{tabular}
\end{table}
 Quantum gravity with Kuz'min or Tomboulis form factor is renormalizable \cite{Kuz89,Tom97,Mod1}; with the string-related profile ${\rm H}^{\rm exp}(\B)=-l^2\B$, it is renormalizable if perturbative expansions with the resummed propagator are allowed \cite{TBM}; with Krasnikov profile ${\rm H}^{\rm exp}(\B)=l^4\B^2$, it is believed to be renormalizable, but the proof is not complete \cite{Kra87}. The role of quadratic curvature operators is to make the theory renormalizable, while the role of the non-local form factors is to preserve unitarity.

The profile ${\rm H}(\B)$ can be defined through the integral
\ba
{\rm H}(\B)&=&\lim_{\s\to 1} {\rm H}_\s(\B)\,,\\
{\rm H}_\s(\B) &:=& \a\int_0^{\s P(\B)} \rmd\omega\, \frac{1-f(\omega)}{\omega}\,,\label{Hz2}
\ea
where $\a>0$ is real, $P(\B)$ is a generic function of $l^2\B$, and $f(\om)$ is arbitrary. The parameter $\s$ is fictitious and has been introduced for later convenience.


\subsection{Non-local dynamics in terms of kernels}

The physics stemming from the action \Eq{action} is a hard nut to crack. Even before quantizing the theory, two fundamental questions arise. How many degrees of freedom are there? Can one solve the dynamics once a finite number of initial conditions are given? While it is easy to see that linear systems have a finite number of d.o.f.\ and of initial conditions, the case with non-linear interactions is highly nontrivial. Directly manipulating the infinitely many derivatives of \Eq{gamma} makes very difficult to answer the above questions.

To put the main result of this Letter in the most direct terms, we trade \Eq{gamma} for kernel functions that obey finite-order differential equations (\emph{master equations}). In general, any operator $\g(\B)$ with finitely or infinitely many derivatives can be written as a non-local kernel function. Consider first the flat-spacetime case \cite{PU}. In momentum space, calling $F$ the Fourier anti-transform of $\g(-k^2)$ (not to be confused with the Fourier transform $\tilde\g(-k^2)$ of $\g(\B)$), for a generic tensor $\vp(x)$ one has
\ba
\hspace{-.7cm}\g(\B)\,\vp(x)\!\!\!\! &=&\!\!\!\! \int\rmd^D k\,\g(-k^2)\,\de^D(k^\mu-\rmi\N^\mu)\,\vp(x)\nonumber\\
								\!\!\!\!&=&\!\!\!\! \int\rmd^D k\,\left[\int\rmd^D z\,F(z)\,\rme^{-\rmi z\cdot k}\right]\,\de^D(k^\mu-\rmi\N^\mu)\,\vp(x)\nonumber\\
								\!\!\!\!&=&\!\!\!\! \int\rmd^D z\,F(z)\,\rme^{z\cdot\N}\vp(x)=\int\rmd^D z\,F(z)\,\vp(x+z)\nonumber\\
							 \!\!\!\!&\stackrel{y:=z+x}{=}&\!\!\!\! \int\rmd^D y\,F(y-x)\,\vp(y)\,.\label{ker}
\ea
Thus, as an operator identity we have
\be\label{Fgam}
F(y-x)=\int\frac{\rmd^Dk}{(2\pi)^D}\,\rme^{\rmi k\cdot (y-x)}\g(-k^2)\,.
\ee
On a curved spacetime, a similar expression holds after generalizing the Fourier transform to an invertible momentum transform where the phases $\exp(\pm\rmi k\cdot z)$ are replaced by the eigenfunctions of the Laplace--Beltrami operator on that spacetime (e.g., \cite{cuta2}). Different operators $\g(\B)$ will lead to different kernels $F$.

We can use this general property of local and non-local operators to write \Eq{action} in terms of kernel functions:
\ba
\hspace{-.7cm}S\!\!\!\!&=&\!\!\!\! \frac{1}{2\kappa^2} \int \rmd^D x \sqrt{-g}\, R + \frac{1}{2\kappa^2} \int \rmd^D x \sqrt{-g(x)}\nonumber\\
\hspace{-.7cm}&&\!\!\!\!\times \int \rmd^D y \sqrt{-g(y)} \int \rmd^D z \sqrt{-g(z)} \, G_{\mu \nu}(x) \nonumber\\
\hspace{-.7cm}&&\!\!\!\!\times \left[\cG(x,y; 1) -[-g(y)]^{-1/2}\de^D(x-y)\right] \,  \tilde{G}(y,z) \, R^{\mu\nu}(z)\,.\label{ExplicitNLG}
\ea
The quantity $\cG$ is the kernel expressing the derivative operator $\exp {\rm H}$. Once a given background metric $g_{\mu\nu}$ is specified, one can calculate the eigenfunctions of the operator $\B$ and write down the generalization of \Eq{Fgam} \cite{cuta2} for $\cG$. However, this procedure is inconvenient because it is background dependent and, in general, the eigenvalue problem of the curved $\B$ can be challenging. Instead, first we define the kernel $\cG$ formally, to understand where it comes from, and then we find a set of master equations which can be solved explicitly to find $\cG$. The formal definition is as a Green function solving
\be\label{Gsol0}
\rme^{-{\rm H}(\B_x)}\cG(x,y;1) = \frac{\de^D(y-x)}{\sqrt{-g(y)}}\,,
\ee
where the number ``1'' in the arguments of $\cG$ will be explained shortly. Similarly, the quantity $\tilde G$ is the Green function solving \cite{Deser2}
\be\label{DW1}
\Box \tilde{G}(y,z) = \frac{\de^D(y-z)}{\sqrt{-g(z)}}\,.
\ee
In this way, $\Box^{-1}$ is expressed in terms of $\tilde G$. By definition, these kernels make \Eq{ExplicitNLG} fully equivalent to \Eq{action}, \emph{if \Eq{Gsol0} and \Eq{DW1} are well defined}. Suppose they are. Then one can formally invert \Eq{Gsol0} and \Eq{DW1} to get
\ba
&\hspace{-.5cm}&\hspace{-1.2cm}\int \rmd^D y \sqrt{-g(y)} \int \rmd^D z \sqrt{-g(z)} \, G_{\mu \nu}(x) \nonumber\\
&\hspace{-.5cm}&\times \left[\cG(x,y; 1) -\frac{\de^D(x-y)}{\sqrt{-g(y)}}\right] \,  \tilde{G}(y,z) \, R^{\mu\nu}(z)\nonumber\\
&\hspace{-.5cm}\stackrel{\textrm{\tiny\Eq{DW1}}}{=}& \int \rmd^D y \sqrt{-g(y)} \int \rmd^D z \, G_{\mu \nu}(x) \nonumber\\
&\hspace{-.5cm}&\times \left[\cG(x,y; 1) -\frac{\de^D(x-y)}{\sqrt{-g(y)}}\right] \,  \de^D(y-z)\frac{1}{\B}\, R^{\mu\nu}(z)\nonumber\\
&\hspace{-.5cm}=& \int \rmd^D y \sqrt{-g(y)} \, G_{\mu \nu}(x) \nonumber\\
&\hspace{-.5cm}&\times \left[\cG(x,y; 1) -\frac{\de^D(x-y)}{\sqrt{-g(y)}}\right] \,\frac{1}{\B}\, R^{\mu\nu}(y)\nonumber\\
&\hspace{-.5cm}\stackrel{\textrm{\tiny\Eq{Gsol0}}}{=}& \int \rmd^D y \, G_{\mu \nu}(x)\left[\rme^{{\rm H}(\B_x)} -1\right]\de^D(x-y) \,\frac{1}{\B}\, R^{\mu\nu}(y)\nonumber\\
&\hspace{-.5cm}=& G_{\mu \nu}(x)\,\g(\B_x)\,R^{\mu\nu}(x)\,.\nonumber
\ea

However, the problem is that \Eq{Gsol0} is not well defined at all! To see this, let us keep the parameter $\s$ in \Eq{Hz2} generic and write \Eq{Gsol} as the $s\to 1$ limit of the formal expression
\be\label{Gsol}
\cG(x,y;\s) = \rme^{{\rm H}_\s(\B_x)} \cG(x,y;0), 
\ee
where $\cG(x,y;0):=[-g(y)]^{-1/2}\de^D(x-y)$. This expression is especially difficult to deal with because in non-local quantum gravities the propagator\footnote{The propagator of this class of theories was amply discussed in the literature; see, e.g., \cite{Mod1,BGKM,TBM,MoRa1,MoRa2,CMN2} and references therein. In particular, we can use the Feynman prescription \cite{Mod1,MoRa2}, which eventually gives rise to a micro-causality violation \cite{CMN2}. The number of d.o.f.\ is not affected by this choice and it will not be mentioned further in this paper.} is suppressed in the ultraviolet, so that its inverse (the form factor $\exp {\rm H}$) explodes in Euclidean momentum space in all realistic cases.\footnote{For instance, for Kuz'min Euclidean form factor (see below) ${\rm H}_\s(-k^2)=\a[\ln(l^2k^2)+\Gamma(0,l^2k^2)+\gamma_{\rm E}]\to+\infty$ as $|k|\to\infty$; for the string form factor, ${\rm H}_\s(-k^2)=\a\s l^2 k^2$; and so on.} To avoid this, we consider the inverse $\cF:=\cG^{-1}$, defined implicitly as
\be\label{defF}
\hspace{-0.5cm}\int \rmd^D z \sqrt{- g(z)} \, \mathcal{G}(x-z; \s) \, \mathcal{F}(z-y; \s) = \frac{\delta^D(x-y) }{\sqrt{-g(y)}} \,,
\ee 
for which the non-local operator in 
\be\label{Fsol1}
\cF(x,y;\s) = \rme^{-{\rm H}_\s(\B_x)} \cF(x,y;0)
\ee 
is damped at high energies. Once $\cF$ is found, one can determine $\cG$ with deconvolution methods \cite{Ulm10}. 


\section{Master equations}

To summarize, we rewrote the non-local action \Eq{action} as \Eq{ExplicitNLG}. Since \Eq{Gsol} is ill-defined at high momenta, we could not find the explicit expression of $\cG$ directly and we had to introduced its inverse $\cF$, defined by \Eq{defF} and obeying \Eq{Fsol1}. Thus, the non-local system \Eq{action} has been fully recast in terms of well-defined kernel functions, equations \Eq{ExplicitNLG} and \Eq{defF}. Now \Eq{Fsol1} is well-defined at high momenta, but it still is a non-local equation with infinitely many derivatives and, in general, we do not know either how to solve it or how to make sense of the initial-value problem, or both.

To address this issue, we make a crucial observation: any form factor can be written in terms of a kernel $\cF$ governed by a simple system of renormalization-group-like equations, which determine how $\cF$ varies in the \emph{space of all possible functionals $P(\B)$}. This space is parametrized by $\s$, where $\s=1$ corresponds to the end of the flow. The exact form of these
master equations depends on the choice of ${\rm H}(\B)$, which is limited by renormalizability and unitarity.

For the general class $f(\om)=\exp(-\om)$ (asymptotically polynomial ${\rm H}^{\rm pol}_\s$; includes Kuz'min and Tomboulis form factors), we can finally write the finite-order differential equations:
\ba
\hspace{-1.5cm}&&\s \partial_\s \cF(x,y;\s) = \a [(\cK \star \cF)(x,y;\s)-\cF(x,y;\s)]\,,\label{Master2}\\
\hspace{-1.5cm}&&(\cK \star \cF)(x,y;\s):= \int \rmd^D x' \sqrt{-g}\, \cK(x,x';\s) \, \cF(x',y;\s)\,,\label{Master3}\\
\hspace{-1.5cm}&& \left[\p_\s+P(\Box_x)\right] \cK(x,y;\s) = 0,\label{Master1}\\
\hspace{-1.5cm}&&\cF(x,y;0) =\cK(x,y;0)= [-g(y)]^{-1/2}\de^D(x-y)\,,\label{Fsol2}
\ea
where $\cK$ is the kernel associated with the operator $f[\s P(\B)]$. This is equivalent to the system \Eq{ExplicitNLG}. In fact, \Eq{Master1} corresponds, in quantum gravities, to diffusion in the correct direction and its solution
\be\label{solmas1}
\cK(x,y;\s) = \rme^{-\s P(\Box_x)} \cK(x,y;0)
\ee
is well defined. Then, noting that
\be
\hspace{-.5cm}\p_\s{\rm H}_\s(\B)=\a P(\B)\,\frac{1-f[\s  P(\B)]}{\s P(\B)}=\a\frac{1-\rme^{-\s P(\B)}}{\s}\,,
\ee
from \Eq{Fsol1} one finds the left-hand side of \Eq{Master2},
\be\nonumber
\s\p_\s\cF= -\s\p_\s{\rm H}_\s(\B)\cF=\a \left[\rme^{-\s P(\B)}-1\right]\cF\,.
\ee
This coincides with the right-hand side of \Eq{Master2}, since
\ba
\hspace{-.7cm}\cK \star \cF\ &\stackbin[\textrm{\tiny \Eq{solmas1}}]{\textrm{\tiny \Eq{Master3}}}{=}& \int \rmd^D x' \sqrt{-g(x')}\, \rme^{-\s P(\B_x)} \cK(x,x';0) \, \cF(x',y;\s)\nonumber\\
&\stackrel{\textrm{\tiny \Eq{Fsol2}}}{=}& \rme^{-\s P(\B_x)}\int \rmd^D x'\,\de^D(x-x') \, \cF(x',y;\s)\nonumber\\
& =& \ \rme^{-\s P(\B)}\cF\,.\nonumber
\ea
Therefore, \Eq{Fsol1} is the solution of \Eq{Master2}.

In the much simpler case of the general class $f(\om)=1-\om$ (exactly monomial ${\rm H}^{\rm exp}_\s$; includes the string-related and Krasnikov exponential form factors), \Eq{Master2}--\Eq{Fsol2} are replaced by just one master equation:
\be\label{MasterSystem2}
[\p_\s+\a P(\Box_x)]\cF(x,y;\s)=0\,,
\ee
whose solution $\cF(x,y;\s) =\rme^{-\a\s P(\B_x)}\cF(x,y;0)$ is well defined and coincides with \Eq{Fsol1}.
 The integro-differential equations \Eq{Master2}--\Eq{Fsol2} or \Eq{MasterSystem2} are the generalizations of the diffusion-equation method \cite{cuta3,cuta7} and of the case with an exactly polynomial ${\rm H}(\B)$ treated in Ref.\ \cite{CMN2}. Thus, the original, formidable non-local problem is reduced to one with a finite number of derivatives where we can count the initial conditions and the field degrees of freedom. Both turn out to be finite.}

\section{Initial conditions for special form factors} 

Let us consider the system \Eq{Master2}--\Eq{Fsol2} with $P(\B) = -l^2\B$. The system given by \Eq{Fsol2}, \Eq{Master2}, 
 and $(\p_\s-l^2\B_x) \cK(x,x';\s) = 0$ is second order in spacetime coordinates and first order in the diffusion parameter $\s$. In synchronous gauge, the metric in $D$ dimensions simplifies to $\rmd s^2 = - \rmd t^2 + h_{i j}(t, {\bf x}) \rmd x^i \rmd x^j$, where $i,j =1, \dots, D-1$, $h_{ij}$ is the metric of the spatial section, and the covariant d'Alembertian operator on a scalar takes the form 
\be
\Box = - \partial_t^2 - \frac{1}{2} h^{ij} \dot{h}_{ij} \partial_t + \frac{1}{\sqrt{h}} \partial_i \left(\sqrt{h} h^{ij} \partial_j \right).
\ee
Therefore, to solve the second-order master equations we only need to specify the spatial metric and its first time derivative. If we insist in having only second-order (in spacetime coordinates) differential equations with a diffusion-like structure, then \emph{there are only two choices for the form factor} \Eq{gamma}: Kuz'min's profile ${\rm H}^{\rm pol}(\B)$ with $P(\B)=-l^2\B$ or the string-related ${\rm H}^{\rm exp}(\B)$ $=-l^2\B$. By removing one ambiguity of the problem [choice of $P(\B)$] once the other [choice of $f(\om)$] is fixed by requiring a diffusion equation, the long-standing question about the uniqueness of non-local gravity is solved. Since the generalized diffusion method \Eq{Master2}--\Eq{Fsol2} can be applied to any $f(\om)$, it permits to classify all allowed form factors.

Therefore, when $P(\B)=-l^2\B$ the whole non-locality in the action and in the equations of motion (EOM) is completely specified by second-order differential equations, together with the metric and the first time derivative of the spatial metric. If we impose the retarded boundary condition $\tilde{G}(x,y) = 0$ for $y^0>x^0$, (\ref{DW1}) defines the retarded Green function $\tilde{G}$ at time $x^0$ only from the value of $h_{ij}$ and its first time derivative, for times $\leq x^0$.

\section{Equations of motion} 

The variation of the non-local action with respect to the metric for a generic form factor is doable but complicated for the nonexpert. While variation of curvature terms gives at most two derivatives, the source of infinite derivatives in the EOM is the variation $\de\g(\B)/\de g^{\mu\nu}$ when $\g$ is expressed (as done in most approaches) as a series of $\B$ powers. There lies the difficulty in understanding the Cauchy problem in non-local quantum gravity. However, thanks to the kernel (instead of series) representation \emph{and} to the master equations \Eq{Master2}--\Eq{Fsol2} or \Eq{MasterSystem2}, we are in a position to count the initial conditions. The strategy is first to calculate the variation of the original system \Eq{action}, where all metric dependence is explicit, and then to use the kernel representation of the resulting non-local derivative operators. On one hand, variation of the curvature terms $R_{\mu\nu}$ and $G_{\mu\nu}$ in the action \Eq{action} gives terms with at most four derivatives acting on the same field:
\be
\frac{(\de G_{\mu\nu})\,\g R^{\mu\nu}+(\de R^{\mu\nu})\,\g G_{\mu\nu}}{\de g^{\mu\nu}} 
\quad\Rightarrow\quad \textrm{4 derivatives}\,,
\ee
where the form factor $\g$ is expressed in terms of the above integral kernels for $\Box^{-1}$ and $\exp {\rm H}$. Notice that, contrary to Deser--Woodard theory \cite{Deser2}, we do not count out two of the four derivatives from the inverse $\B$ represented by $\tilde G$.\footnote{In fact, the role of the $\B^{-1}$ operator is different in the two theories. In the present case, it cancels the $O(\B)$ leading term in the numerator of the form factor \Eq{gamma}, making it entire: in a derivative expansion, $\g(\B)={\rm const}+O(\B)$. See \cite[section 3.3]{CMN2} for more details.} The variation of $\g$ is
\ba
\hspace{-.4cm}\delta \g &=& -\frac{1}{\B}(\de\B)\g+\frac{\a l^{2n}}{\B}\sum_{k=1}^n\int_0^1\!\!\rmd s\rme^{s H_\s}\!\!\!\int_0^\s\!\rmd\s'\!\!
\int_0^1\rmd s'\nonumber\\
&&\times\rme^{-s\s'P}\Box^{k-1}(\delta \Box)\Box^{n-k} \rme^{(s-1)\s'P}\rme^{(1-s)H_\s}.\label{finalGammaVar}
\ea
It involves form factors of the type $\B^k$ with $k>0$, $\Box^{-1}$, $\exp \Box^n$ (for a monomial $P(\B)=(l^2\B)^n$), and $\exp {\rm H}$. The operators $\B^k$ are left as they are, since their kernel representation is the $2k$-th derivative of the Dirac delta and it does not reduce the number of derivative operators in the equations. All the other operators admit the above kernel representations, satisfying \Eq{DW1} in the case of $\B^{-1}$,\footnote{This operator should be treated in the kernel representation rather than trying to absorb it with positive powers of the $\B$ to combine it to an operator $\B^{k-2}\de\B\dots$. This is clear from the $k=1$ term in \Eq{finalGammaVar}, which is of the irreducible form $\B^{-1}\de\B\,\B^{n-1}$ (recall that $\B$ and $\de\B$ do not commute).} \Eq{Fsol2} and \Eq{MasterSystem2} in the case of $\exp\B^n$, and the master equations \Eq{Master2}--\Eq{Fsol2} in the case of $\exp {\rm H}$. Therefore, using the kernel representations \Eq{finalGammaVar} contributes with $2(k-1)+2+2(n-k)=2n$ derivatives, which act on the two derivatives in the Ricci or Einstein tensor producing terms with at most $2n+2$ derivatives acting on the same field:
\be
\frac{G_{\mu\nu}\,(\de\g) R^{\mu\nu}}{\de g^{\mu\nu}} 
\quad\Rightarrow\quad \textrm{$2n+2$ derivatives}\,.
\ee
Thus, the EOM with respect to the metric are of order $2n+2$. This is one of the main findings of the present work: that there exist suitable kernel representations of the form factors obeying finite-order equations.
 The derivative order of the system given by the EOM and the master equations is thus $2n+2$.

The conclusion is that the EOM can be expressed in terms of the kernels $\cK$ and $\cG$ living in the space of form factors and the Green function $\tilde{G}(x,y)$. These kernels are non-local because they depend on two spacetime points, but they are determined by a set of equations independent of the actual EOM. This set may be more or less difficult to solve, but it is well defined. Therefore, the EOM coming from \Eq{ExplicitNLG} are non-local integral equations but \emph{finite} differential equations: for $P(\B)=O(\B^n)$, they are of derivative order $2n+2$ and we must specify $2n+2$ initial conditions. This result agrees with the counting in a non-local scalar field theory \cite{cuta3,Tom15}, and with the diffusion method in Lagrangian formalism \cite{CMN2}. Deser--Woodard non-local gravity \cite{Deser2} admits a well defined counting, too, although the diffusion method is not needed there due to the different nature of the form factor therein.

Up to this point, we concentrated on a class of super-renor\-mal\-izable theories, but the main result is insensitive to the introduction of other special operators that make the theory finite in even dimension. Here by special we mean non-local ``killer operators'' \cite{MoRa1} at least cubic in the curvature, namely, $\mathcal{R} \gamma_{{\rm k},1}(\Box) \mathcal{R}^2$, $\mathcal{R}^2 \gamma_{{\rm k},2}(\Box) \mathcal{R}^2$, $\ldots$, $\mathcal{R}^{D/2} \gamma_{{\rm k},\cdots}(\Box) \mathcal{R}^2$ and $\mathcal{R} \,\Box \,\mathcal{R}^{\frac{D}{2}-2}$ $\times\gamma_{{\rm k},\cdots}(\Box) \mathcal{R}^2$, where $\gamma_{{\rm k},\cdots}$ may all differ from one another and, again, $\mathcal{R}$ is the generalized curvature. These operators increase the order of the EOM. In $D=4$, killer operators are cubic or quartic in the Riemann tensor and the order of the EOM is the same, namely, $2n+2$. In particular, for $n=1$ four-dimensional finite non-local gravity only needs four initial condition at the non-perturbative level.

\section{Degrees of freedom} 

Having counted the number of initial conditions, we also comment on the number of field degrees of freedom, i.e., the number of independent components of the tensor fields populating the theory. We introduce two auxiliary fields, a rank-2 symmetric tensor $\phi^{\mu\nu}$ and a scalar field $\psi$, to infer the exact number of d.o.f. Let us consider the Lagrangian
\ba
2\kappa^2\cL&=& R+2 G_{\mu \nu}\,\gamma(\Box)\,\phi^{\mu\nu}-\phi_{\mu\nu}\,\gamma(\Box)\,\phi^{\mu\nu}\nonumber\\
&&+R\,\gamma(\Box)\,\psi+\psi\,\gamma(\Box)\,\psi/(D-2)\,.
\label{phih}
\ea
The EOM for the tensor $\phi^{\mu\nu}$ and the scalar $\psi$ are
\be
\phi_{\mu\nu} = G_{\mu\nu}\,,\qquad \psi=\phi=G\,.\label{EOMAUX}
\ee
$\psi$ is just the trace of $\phi_{\mu\nu}$ and is not an independent degree of freedom \cite{CMN2}. Eliminating the auxiliary fields from the action, we end up with the original action (\ref{action}). Notice that (\ref{EOMAUX}) implies the transverse condition $\nabla^\mu \phi_{\mu\nu}=0$ on-shell, as a consequence of Bianchi identity. The EOM for the metric are more involved, but not overly so, and agree with the case without auxiliary fields \cite{CMN2}. It turns out that we only deal with second-order differential equations \cite{CMN2}, up to $\g$ factors that, as we have shown above, can be dealt with the diffusion-equation method without increasing the derivative order. On the other hand, by a simple count of the independent components of the fields, we find that the d.o.f.\ are:
\begin{enumerate}
\item[(I)] Graviton $g_{\mu\nu}$: symmetric $D\times D$ matrix with $D(D+1)/2$ independent entries, to which one subtracts $D$ Bianchi identities $\N^\mu G_{\mu\nu}=0$ and $D$ diffeomorphisms (the theory is fully diffeomorphism invariant). Total: $D(D-3)/2$. In $D=4$, there are 2 degrees of freedom, the usual polarization modes. 
\item[(II)] Tensor $\phi_{\mu\nu}$: symmetric $D\times D$ {matrix to} which one subtracts $D$ transverse conditions $\N^\mu \phi_{\mu\nu}=0$. Total: $D(D-1)/2$. In $D=4$, there are 6 degrees of freedom.
\end{enumerate}
The grand total is $D(D-2)$. In the minimal case $n=1$ in $D=4$, the EOM are fourth order and the degrees of freedom are eight, just like in local Stelle quadratic gravity \cite{Ste77}. In arbitrary $D$ dimensions, the EOM are still fourth order, but the number of degrees of freedom is $D(D-2)$. Therefore, in general the number of d.o.f.\ is not half the number of initial conditions, as in local gravity, except in four dimensions. This $D=4$ case is only a coincidence. On one hand, the counting of field d.o.f.\ (number of independent field components) is the same as in Stelle gravity because it is not affected by the presence of the form factor. On the other hand, as we saw, the number of initial conditions depends on the choice of form factor but is independent of the number of dimensions.

We here expand on the nature of these particle d.o.f.\ in Minkowski spacetime, following closely \cite{Ste78}. First we focus on the theory in $D=4$ dimensions.

In order to distinguish the d.o.f.\ of the theory (\ref{action}), it is useful to separate the massive from the massless fields. We consider the action (\ref{action}) at the quadratic order in the perturbation $h_{\mu\nu}$, the latter being defined by $g_{\mu\nu} = \eta_{\mu\nu} + 2 \kappa h_{\mu\nu}$:
%
\be
\hspace{-0.5cm}
\mathcal{L}^{(2)} = - \frac{1}{2} h^{\mu\nu}  \Box \, \rme^{\rm H}  P^{(2)}_{\mu\nu \rho \sigma}  h^{\rho \sigma}  -  h^{\mu\nu} \Box \, \rme^{\rm H}  P^{(0)}_{\mu\nu \rho \sigma}  h^{\rho \sigma},
\label{second order}
\ee
 where the projectors $P^{(2)}$ and $P^{(0)}$ are defined as
\ba
P^{(2)}_{\mu\nu\rho\sigma} &=& \frac{1}{2} \left( \theta_{\mu\rho} \theta_{\nu\sigma} + \theta_{\mu\sigma} \theta_{\nu\rho} \right) - P^{(0)}_{\mu\nu\rho\sigma} \, , \nonumber \\
P^{(0)}_{\mu\nu\rho\sigma} &=& \frac{1}{3}  \theta_{\mu\nu} \theta_{\rho\sigma} \, , \quad  \theta_{\mu\nu} = \eta_{\mu\nu} - \frac{\partial_\mu \partial_\nu}{\Box} \, .\nonumber
\ea
We now proceed as in local quadratic gravity \cite{Ste78} and, in (\ref{second order}), we first replace the graviton $h_{\mu\nu}$ with $h_{\mu\nu} + \Sigma_{\mu\nu}$ and, after some intermediate field redefinitions, the tensor $\Sigma_{\mu\nu}$ with 
\be 
\Sigma_{\mu\nu} = \Psi_{\mu\nu} + \eta_{\mu\nu} \, \chi - 2 \frac{\rme^{\rm H} -1}{\Box} \partial_\mu \partial_\nu \chi \,,
\ee
where $\Psi_{\mu\nu}$ is a symmetric rank-2 tensor and $\chi$ is a scalar. The outcome at the second order in the perturbations is the following Lagrangian:
\ba
\mathcal{L}^{(2)} &=& \mathcal{L}_{\rm E} (h_{\mu\nu})
- 3 \partial_\mu \chi \partial^\mu \chi + 3 \chi \frac{\Box}{\rme^{\rm H} - 1} \chi- \mathcal{L}_{\rm E} (\Psi_{\mu\nu}) 
 \nonumber \\
&&- \frac{1}{2} \Psi_{\mu\nu}     \frac{\Box}{\rme^{\rm H} - 1}   \Psi_{\mu\nu} 
+  \frac{1}{2} \Psi^{\rho}_\rho    \frac{\Box}{\rme^{\rm H} - 1}    \Psi^{\sigma}_\sigma 
\,,
\label{second order2}
\ea
where $\B=\p_\mu\p^\mu$ and we introduced Einstein's linearized Lagrangian (for a generic field $Z_{\mu\nu}$)
\ba
\mathcal{L}_{\rm E} (Z_{\mu\nu}) &=& \frac{1}{2} Z_{\mu\nu} \Box Z^{\mu \nu} 
- \frac{1}{2} Z^\rho_\rho \Box Z^\sigma_\sigma+ Z^{\mu\nu} \partial_\mu \partial_\nu Z^\lambda_\lambda\nonumber\\  
&&-  Z^{\mu\nu} \partial_\rho \partial_\nu Z^\rho_\mu \, . 
\ea
Selecting the gauge-independent terms of the action, we get
\ba
\mathcal{L}^{(2)} &=& \mathcal{L}_{\rm E} (h_{\mu\nu}) 
+ 3 \chi  \frac{\Box}{1 - \rme^{-{\rm H}}} \chi\nonumber\\
&&- \frac{1}{2}  \Psi_{\mu\nu} \left(\eta^{\mu\rho} \eta^{\nu\sigma}- \eta^{\mu\nu} \eta^{\rho\sigma} \right) \frac{\Box}{1 - \rme^{-{\rm H}}} \Psi_{\rho \sigma} .
\label{second orderF}
\ea
Notice that the local limit of the form factors in \Eq{second orderF} is
\be\label{ffsto}
\g^{-1}\,\rme^{{\rm H}}=\frac{\Box}{1-\rme^{-{\rm H}}}\simeq \frac{\Box}{{\rm H}},
\ee
which reduces to Stelle gravity (mass terms for $\chi$ and $\psi_{\mu\nu}$) only in the theory with string-related form factor, where ${\rm H}\propto \B$. In any other case, the local limit of the theory is not Stelle gravity.

The spin structure of the fields is not affected by the form factors, which are entire functions, and is the same as in Stelle gravity. Therefore, we do not need to repeat the discussion in \cite{Ste77,Ste78,HOW}. However, non-locality radically changes the propagation of these fields and, ultimately, the physical content of the theory. Thus, on one hand the theory (\ref{action}) describes a massless graviton field, a spin-two field\footnote{Taking the divergence and the trace of the equations of motion (as done in \cite{HOW} for Stelle gravity) one obtains two conditions $\N^\mu\Psi_{\mu\nu}=0$ and $\Psi_\mu^\mu=0$ on Minkowski spacetime. The field $\Psi_{\mu\nu}$ is symmetric, transverse, and traceless and is, therefore, an eigenstate of the spin Casimir operator $S^2$ with eigenvalue 2.} with kinetic term with the wrong sign (analogous to Stelle's Pauli--Fierz massive ghost field), and a scalar field. On the other hand, the gauge-invariant terms of the propagators for $\chi$ and $\Psi_{\mu\nu}$ are both proportional to the inverse of \Eq{ffsto}, $\g\,\rme^{-{\rm H}}$, which has no poles by definition of ${\rm H}$. The conclusion is that the spin-2 ghost and the scalar present in Stelle local theory \cite{Ste77} do not propagate at the perturbative level in non-local gravity. 

Furthermore, the six non-propagating d.o.f.\ $\Psi_{\mu\nu}$ and $\chi$ are exactly the same of $\phi_{\mu\nu}$ up to a change of basis. We can safely conclude that the action (\ref{phih}) describes exactly the same d.o.f.\ of Stelle's theory, that are now harmless thanks to the non-locality of the action. Finally, it was recently proved \cite{BrMo} that Minkowski spacetime is stable not only at higher perturbative order, but also to all orders in the graviton perturbation. This result was then extended to any Ricci-flat spacetime (stable in non-local gravity if stable in general relativity) \cite{BCM}. Therefore, the field $\phi_{\mu\nu}$, or the fields $\chi$ and $\Psi_{\mu\nu}$, never propagate at any arbitrarily high perturbative order.

In $D$ dimensions, the number of independent components or the fields changes, but the spin decomposition into a graviton, a spin-2 ghost and a scalar is the same. This was showed in \cite{AAM} for higher-derivative local theories, but it is true also for non-local theories, since the form factors only affect the propagation of these modes, not their spin. The non-propagation of the spin-2 and the scalar d.o.f.\ can be checked by coupling the graviton to the most general energy tensor $\Theta^{\mu\nu}$ and computing the transition amplitude in momentum space, namely,
\be
\mathcal{A} :=\Theta^{\mu\nu}(k) \mathcal{O}^{-1}_{\mu\nu, \rho \sigma} \Theta^{\rho \sigma}(k) \, .
\label{ampli1b}
\ee
where $\mathcal{O}^{-1}_{\mu\nu, \rho \sigma}$ is the propagator for the theory (\ref{action}). The result is
\ba
\hspace{-0.7cm}
\mathcal{A} &=& \frac{\Theta_{\mu\nu} \Theta^{\mu\nu} -\frac{(\Theta^\mu_\mu)^2}{D-2}}{k^2}- \left[ \Theta_{\mu\nu} \Theta^{\mu\nu} -\frac{ (\Theta^\mu_\mu)^2 }{D-1} \right] \frac{1 - \rme^{-{\rm H}}}{k^2} \nonumber\\
&& + \left[ \frac{(\Theta^\mu_\mu)^2}{(D-1)(D-2)} \right] \frac{1 - \rme^{-{\rm H}}}{k^2} \, . 
\ea
Again, the graviton is the only one propagating in any dimension, while for the massive spin-2 field and the scalar we do not have any pole, since $[1-\exp(-{\rm H})]/k^2$ is entire.

We have proved that the number of physical d.o.f.\ is at most 8 in $D=4$ dimensions and in the minimal theory, but at the moment we cannot exclude the possibility that other constraints, identities, or symmetries will further reduce this number. However, our result already places strong stakes on the theory. At high energy and short distances, the equivalence principle states that spacetime is Minkowski. The extra d.o.f.\ might propagate on other, non-Ricci-flat spacetimes (which, if the extra perturbations have a ghost-like or tachyonic nature, can be unstable and then decay instantaneously, or in a finite time since the theory is non-local \cite{GaVi,Sbi14,JMM}, into another spacetime) but not in the ultraviolet regime. The literature is replete with other examples where some degrees of freedom are non-physical in certain backgrounds. In Ho\v{r}ava--Lifshitz gravity, some scalar modes do not propagate at the linear perturbation level on a cosmological background due to the absence of time derivatives in their equations of motion \cite{KLM,GWBR,BPS}. Certain quantization prescriptions of Lee--Wick gravity theories have modes, called Lee--Wick particles \cite{MoSh} or fakeons \cite{Ans18}, that propagate inside Feynman diagrams but decouple from the physical spectrum on Minkowski spacetime. Our results are more general because they hold at arbitrary perturbative order and for a wide class of backgrounds, but they do not differ from other situations like these where not all the fields in the theory appear in the physical spectrum on special backgrounds (in our case, all Ricci-flat spacetimes are ghost-free). The extra modes propagate inside Feynman loops (off-shell) but not in external legs (on-shell).

\section{Applications and conclusions} 

Some years ago, the diffusion-equation method revealed that a string-motivated non-local scalar field theory with exponential operators has a finite number of initial conditions and non-perturbative degrees of freedom \cite{cuta3}. In this Letter, we converted the infinite number of derivatives of renormalizable non-local gravities with highly nontrivial non-local operators into integral kernels living in the space of all possible form factors and that do not carry pathological non-locality. The main action \Eq{ExplicitNLG} must be placed side by side with the finite-order differential equations given by the system \Eq{Master2}--\Eq{Fsol2} or \Eq{MasterSystem2}. The system of EOM for the metric and the kernels is of finite order $2n+2$ for the metric, for a form factor \Eq{gamma} with ${\rm H}$ given by ${\rm H}^{\rm pol}$ with $P(\B)\propto\B^n$. For the special form factor with $n=1$, where kernels satisfy second-order differential equation, the EOM for the metric are fourth order and the number of non-perturbative degrees of freedom is eight in four dimensions. Two of these d.o.f.\ are perturbative and well known in the literature and, thanks to asymptotic freedom, they describe the theory completely at small scales on a Minkowski background \cite{CaMo2}, i.e., at the scales of the local inertial frame of the observer where the background is approximately flat. The new degrees of freedom found in this paper are non-perturbative and might play a capital role at large scales but not at high energy. 
By large scales, one means scales where tidal forces become important. The diffusion method proposed here reaches those solutions that cannot be found with the available methods applied to linearized EOM. 
This is the most direct and important consequence of our findings for the physics of non-local gravity.

These results repair the old and bad reputation of non-local theories for having an ill-defined Cauchy problem. Armed with the generalized diffusion method, one can construct unambiguous analytic solutions of cosmological and astrophysical backgrounds, previously inaccessible via the typical \emph{Ansatz} $\B R=\la R$ in the literature. From these new, fully non-perturbative solutions, one will be able to extract conclusions on the stability or absence of big-bang and black-hole singularities, find predictions on the cosmic evolution induced by non-local gravity, and check the theory at the non-perturbative level against present and near-future observations, such as those of \textsc{Planck} \cite{P186,P1810} and LIGO \cite{Ab16a,Ab16b}, according to a varied battery of phenomenological tests that has already been effective for string cosmology and other quantum gravities \cite{CQC,revmu}. Supported by these, other recent, and upcoming data, the phenomenology and testing of non-local quantum gravity is about to bloom.


\section*{Acknowledgments} 

\noindent G.C.\ and L.M.\ are supported by the I+D grant FIS2017-86497-C2-2-P of the Spanish Ministry of Science, Innovation and Universities.



\end{document}